\documentclass[aps,prl,10pt,onecolumn,superscriptaddress,nobibnotes]{revtex4-2}
\usepackage{amssymb}
\usepackage{epsfig}
\usepackage{amsmath}
\usepackage{times}
\usepackage{booktabs}
\usepackage{array}

\usepackage{etoolbox}
\newcommand{\lastbiblabel}{LastBibItem}
\makeatletter
\patchcmd{\NAT@bibsetnum}{LastBibItem}{\lastbiblabel}{}{\errmessage{Bibliography width patch failed}}
\patchcmd{\endthebibliography}{LastBibItem}{\lastbiblabel}{}{\errmessage{Bibliography label patch failed}}
\let\auto@bib@innerbib\@empty
\makeatother

\newcommand{\paperpdfinfo}{%
  /Author (Jiaqi Nie; Xincheng Shu; Yong Min; Qi Xuan; Zhongyuan Ruan)
  /Title (Core-periphery structure induces double phase transitions in information cascades)
}
\AtBeginDocument{%
  \ifdefined\pdfinfo
    \pdfinfo{\paperpdfinfo}%
  \else
    \ifdefined\XeTeXversion
    \fi
  \fi
}

\begin{document}
	
\title{Core-periphery structure induces double phase transitions in information cascades}

\author{Jiaqi Nie}
\thanks{These authors contributed equally to this work.}
\affiliation{Institute of Cyberspace Security, Zhejiang University of Technology,  Hangzhou 310023, China}
\affiliation{Binjiang Institute of Artificial Intelligence, Zhejiang University of Technology, Hangzhou 310056, China}
\affiliation{Department of Network and Data Science, Central European University, Vienna, 1100, Austria}

\author{Xincheng Shu}
\thanks{These authors contributed equally to this work.}
\affiliation{Computational Communication Research Center, Beijing Normal University, Zhuhai, 519087, China}
\affiliation{School of Journalism and Communication, Beijing Normal University, Beijing, 100875, China}

\author{Yong Min}
\affiliation{Computational Communication Research Center, Beijing Normal University, Zhuhai, 519087, China}
\affiliation{School of Journalism and Communication, Beijing Normal University, Beijing, 100875, China}

\author{Qi Xuan}
\affiliation{Institute of Cyberspace Security, Zhejiang University of Technology,  Hangzhou 310023, China}
\affiliation{Binjiang Institute of Artificial Intelligence, Zhejiang University of Technology, Hangzhou 310056, China}

\author{Zhongyuan Ruan}
\email{zyuan.ruan@gmail.com}
\affiliation{Institute of Cyberspace Security, Zhejiang University of Technology,  Hangzhou 310023, China}

\begin{abstract}
Double phase transitions in spreading processes are commonly associated with additional dynamical or structural complexity. Here, we show that such transitions can emerge naturally in prevalent networks exhibiting a core-periphery (CP) organization. Specifically, we consider threshold-driven spreading on CP networks constructed from the stochastic block model, which is characterized by three parameters: the relative core size $r$, the CP coupling strength $\mu$, and the fraction of intra-block link allocation $\alpha$. We find that, by varying the allocation of links between the core and periphery, the system can generate a global cascade window in the parameter spaces, bounded by distinct phase transitions. At the upper boundary, where the periphery becomes sparse, the interplay between the structural inertia (resistance to state changes) of the core and the spreading constraints of the periphery can produce a double phase transition via a two-stage cascading mechanism. This double transition may change type from discontinuous-continuous to discontinuous-discontinuous as the model parameters vary in the limit of a vanishingly small initial seed. Our results reveal a simple structural route to double phase transitions of distinct types and highlight the role of CP structure in shaping threshold-driven spreading dynamics. 
\end{abstract}

\maketitle

Many real-world networks exhibit a core-periphery (CP) structure, in which a subset of nodes are densely interconnected, whereas the remaining nodes are loosely connected \cite{Rombach2017}. Unlike community structures, where links between different blocks are typically sparse \cite{Newman2006,Nematzadeh2014}, nodes in the core can maintain strong connections with peripheral nodes. Such a structural feature has been widely observed in a variety of complex systems, including social, transportation, biological and economic networks \cite{Verma2014,Fagiolo2010,Masuda2006,Lee2014,Anastasiou2019}. A notable example is the world airline network, which consists of a small, almost fully connected core and a vast, approximately tree-like periphery \cite{Verma2014}. Similarly, social networks such as Twitter often contain an elite group that forms a small but densely connected core of influential individuals \cite{Masuda2006}. 

Considerable attention has been devoted to understanding the emergence of CP structures and to developing methods for their detection in real-world networks \cite{Borgatti2000,Zhang2015,Rossa2013,Holme2005,Battiston2018,Nie2025,Bergermann2025,Verma2016,Colomer2014}. Beyond these efforts, understanding how CP architecture shapes dynamical processes, such as information spreading, is another important problem in network science. Previous studies have investigated the influence of the core on spreading dynamics, with a particular focus on simple contagion processes (e.g., epidemic spreading models) \cite{Kitsak2010,Malliaros2016,Nadini2020,Huang2019}, where interactions between pairs of nodes are independent. It is generally believed that the densely connected core plays a crucial role in facilitating large-scale information diffusion. 

However, real-world information spreading often involves more complex mechanisms like social reinforcement, where the adoption behavior depends on the states of all neighbors and is naturally described by threshold-based models \cite{Granovetter1983,Karsai2014,Ruan2015}. The interplay between threshold dynamics and network structure has been extensively studied \cite{Watts2002,Payne2009,Yagan2012,Galstyan2007}. In random networks, for instance, Watts showed that the phase diagram in the $(\phi, z)$ space (where $\phi$ is the average threshold of nodes and $z$ is the average degree of the underlying network) exhibits a global cascade window bounded by a continuous transition at low $z$ and a discontinuous transition at high $z$ \cite{Watts2002}. Nevertheless, a systematic understanding of how CP structures alter the phase diagram and transition behavior of complex contagion remains an open question, despite several studies investigating threshold-based models on CP networks from different perspectives \cite{Curato2016,Fu2025}.

In this work, we adopt the linear threshold model to study how CP structure influences phase transitions in information cascades. We show that  redistributing links between the core and periphery can drive the system through distinct phase transitions. Notably, at the upper boundary of the global cascade regime, where the periphery becomes sparse, a double phase transition may appear due to the interplay between the structural inertia of the core and the structural constraints of the periphery. Previous studies of information spreading have associated such double transitions with heterogeneous dynamical rules (e.g., simple versus complex contagions) \cite{Min2018,Shu2024}, asymmetric multilayer couplings \cite{Yang2023}, nontrivial network geometry \cite{Gao2015}, or group-based interactions \cite{Wang2018}. Here, we show that double phase transitions can arise from a particularly simple and prevalent form of mesoscale organization, namely, a CP structure, without assuming additional dynamical or structural complexity. Moreover, we find that the nature of the double transition may change with the model parameters, a feature not captured by previous models. 

\begin{figure}[tbp]
\centering
\includegraphics[width=0.65\linewidth]{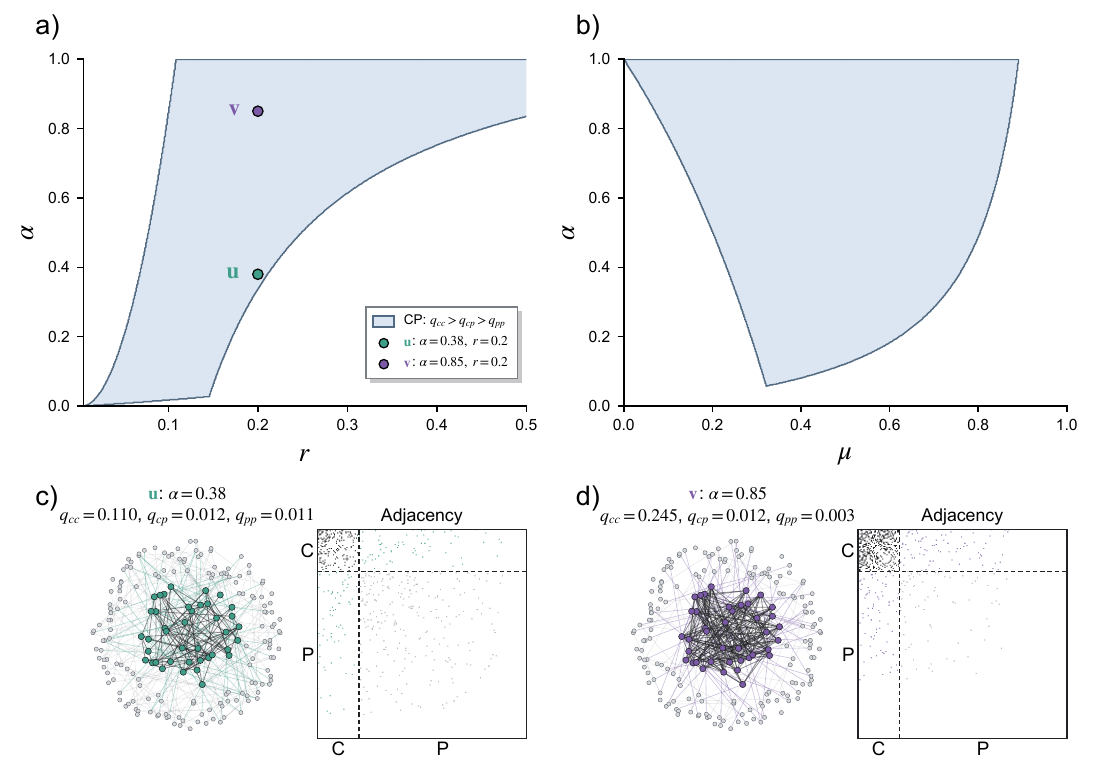}
\caption{(color online). Illustration of the CP region in the parameter space for (a) $(r,\alpha)$ with $\mu=0.25$ and (b) $(\mu,\alpha)$ with $r=0.2$. Panels (c) and (d) show the corresponding network structures and adjacency matrices for two representative points $u=(0.2,0.38)$ and $v=(0.2,0.85)$ in the $(r,\alpha)$ plane, respectively. Here, $N=200$ and $E=300$.}
\label{Fig:1}
\end{figure}

We begin by constructing a CP network based on the stochastic block model. Consider a set of $N$ nodes and $E$ undirected links. The $N$ nodes are first partitioned into two blocks: one containing $rN$ nodes and the other containing $(1-r)N$ nodes.  Among the $E$ links, a fraction $\mu E$ is randomly placed between the two blocks, $\alpha (1-\mu)E$ within the first block, and the remaining within the second block. The parameters $r$, $\mu$ and $\alpha$ are constrained to $[0,1]$. Without loss of generality, we refer to the first block as the core, and the second as the periphery. In most real-world networks, the core is typically smaller than the periphery. Accordingly, we assume $r\le 0.5$ throughout the paper. For a CP structure, the connection probabilities between nodes within the core ($q_{cc}$), between the core and periphery ($q_{cp}$ or $q_{pc}$), and within the periphery ($q_{pp}$) satisfy $q_{cc}>q_{cp}>q_{pp}$, where 
\begin{eqnarray}\label{eq:2}
q_{cc}&=&\frac{2(1-\mu)\alpha E}{N_c(N_c-1)}, \nonumber\\
q_{cp}&=&q_{pc}=\frac{\mu E}{N_c N_p}, \nonumber\\
q_{pp}&=&\frac{2(1-\mu)(1-\alpha) E}{N_p(N_p-1)},
\end{eqnarray}
with $N_c= rN$ and $N_p= (1-r)N$. The three parameters $\mu$, $r$ and $\alpha$ have distinct structural roles: $\mu$ controls the coupling strength between the core and the periphery, $r$ determines the relative size of the core (and equivalently, the periphery), and $\alpha$ controls the allocation of intra-block links between the two blocks --- increasing $\alpha$ produces a denser core and a sparser periphery, and hence a more pronounced CP structure [Fig. \ref{Fig:1} (c) and (d)]. These three parameters jointly define a CP region in the three-dimensional parameter space, in which the inequality $q_{cc}>q_{cp}>q_{pp}$ holds. For convenience, one of these parameters can be fixed, and the CP region can be visualized through a two-dimensional cross section, as illustrated in Fig. \ref{Fig:1} (a) and (b). 

The degree distribution of the core $P_c(k)$ follows a Poisson distribution with mean $z_c=z_{c}^{in}+z_{c}^{out}$, where $z_{c}^{in}=\frac{2(1-\mu)\alpha E}{rN}$ is the average number of intra-core links per node, and $z_{c}^{out}=\frac{\mu E}{rN}$ is the average number of links connecting a core node to the periphery. Similarly, the degree distribution of the periphery, $P_p(k)$, follows a Poisson distribution with mean $z_p=z_{p}^{in}+z_{p}^{out}$, where $z_{p}^{in}=\frac{2(1-\mu)(1-\alpha)E}{(1-r)N}$ and $z_{p}^{out}=\frac{\mu E}{(1-r)N}$ are the average numbers of intra- and inter-block links per periphery node, respectively. 

For information spreading dynamics, we use the linear threshold model. Each node can be in either an active or inactive state. At time $t=0$, a small fraction $\rho_0$ of nodes is randomly initialized as active, while all remaining nodes are inactive (here we consider a single initial seed, i.e., $\rho_0=1/N$; larger initial sizes are considered in the Supplemental Material). We assume that the initial seeds are all located in the periphery. At time $t$, each inactive node updates its state according to the threshold condition: it becomes active if the fraction of its active neighbors exceeds a predefined threshold $\phi$ (i.e., $m/k \ge \phi$, where $k$ is node degree and $m$ is the number of active neighbors, respectively). Once activated, nodes remain active for the rest of the dynamics. The dynamics ends when no further state changes occur. 

This model can be studied analytically by the tree-like approximation \cite{Gleeson2007,Gleeson2008}. The main idea is to approximate the original network (in the limit of $N\to \infty$) as an infinite-depth tree and assume that a node at level $n$ can only be influenced by nodes at level $n-1$ (the tree is labeled from the bottom, $n = 0$, to the top, $n =\infty$). We define $y_n^c$ ($y_n^p$) as the probability that a randomly selected core (periphery) node at level $n$ being active, given that its parent node at level $n+1$ is inactive (so that it can only be influenced by nodes at level $n-1$). The quantities $y_n^c$ and $y_n^p$ satisfy the recursive relations
\begin{eqnarray}\label{eq:yn-yp}
y_{n+1}^{c}&=&\rho_0^{c}+(1-\rho_0^{c})F(y_n^c,y_n^p), \label{eq:yn-yp1}\\ 
y_{n+1}^{p}&=&\rho_0^{p}+(1-\rho_0^{p})G(y_n^c,y_n^p), \label{eq:yn-yp2}
\end{eqnarray}
where $\rho_0^{c(p)}$ denotes the initial fraction of active nodes in the core (periphery). In our model, we have $\rho_0^{c}=y_0^c=0$ and $\rho_0^{p}=y_0^p=\frac{\rho_0}{1-r}$. The functions $F(y_n^c,y_n^p)$ and $G(y_n^c,y_n^p)$, which represent the probabilities that an inactive core or periphery node becomes active due to the influence of its child neighbors, are given by 
\begin{eqnarray}\label{eq:F-G}
F&=&\sum_k \frac{kP_{c}(k)}{z_c} \sum_{m=\lceil \phi k\rceil}^{k-1}  \begin{pmatrix} k-1\\  m \end{pmatrix} (\bar y_n^{c})^m (1-\bar y_n^{c})^{k-1-m}, \nonumber\\ 
G&=&\sum_k \frac{kP_{p}(k)}{z_p} \sum_{m=\lceil \phi k\rceil}^{k-1}  \begin{pmatrix} k-1\\  m \end{pmatrix} (\bar y_n^{p})^m (1-\bar y_n^{p})^{k-1-m}, \nonumber
\end{eqnarray}
where the term $\frac{kP_{c(p)}(k)}{z_{c(p)}}$ represents the degree distribution of core (periphery) nodes in the tree, corresponding to the distribution of the number of nearest neighbors in the original network. $\bar y_n^{c(p)}$ means the probability that a randomly chosen child neighbor of a core (periphery) node at level $n$ being active, which can be written as (see the Supplemental Material)
\begin{eqnarray}\label{eq:inter-yn}
\renewcommand{\arraystretch}{1.5}
 \begin{pmatrix} \bar y_n^{c}\\ \bar y_n^{p} \end{pmatrix} = \begin{pmatrix} \frac{2(1-\mu)\alpha}{2(1-\mu)\alpha+\mu} & \frac{\mu}{2(1-\mu)\alpha+\mu}\\  \frac{\mu}{2(1-\mu)(1-\alpha)+\mu}  & \frac{2(1-\mu)(1-\alpha)}{2(1-\mu)(1-\alpha)+\mu} \end{pmatrix} \begin{pmatrix} y_n^{c}\\  y_n^{p} \end{pmatrix} .
\end{eqnarray}

The average normalized size of global cascades in the core (periphery) at the steady state is given by
\begin{eqnarray}\label{eq:frac_core}
\rho_\infty^{c(p)}&=&\rho_0^{c(p)}+(1-\rho_0^{c(p)})\sum_k P_{c(p)}(k) \nonumber \\ 
&\times& \sum_{m=\lceil \phi k\rceil}^{k}  \begin{pmatrix} k\\  m \end{pmatrix} (\bar{y}_\infty^{c(p)})^m (1-\bar{y}_\infty^{c(p)})^{k-m}.
\end{eqnarray}
Consequently, the overall average normalized size of global cascades in the entire network is $\rho_\infty=r\rho_\infty^c+(1-r)\rho_\infty^p$.

\begin{figure}[tbp]
\centering
\includegraphics[width=0.80\linewidth]{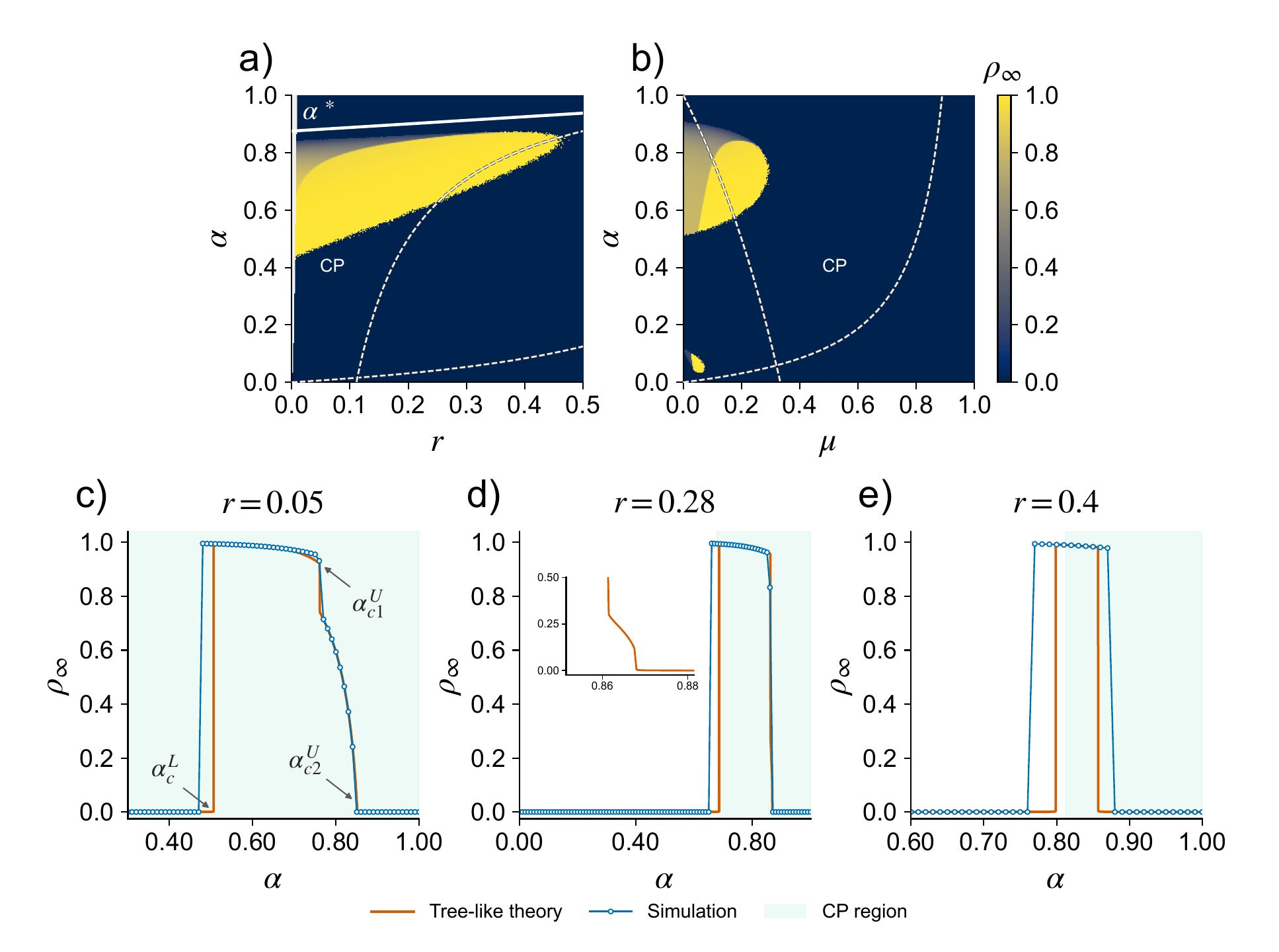}
\caption{(color online). Average normalized size of global cascades at the steady state, $\rho_\infty$, as a function of (a) $r$ and $\alpha$ with $\mu=0.2$, and (b) $\mu$ and $\alpha$ with $r=0.2$. The dashed lines correspond to the solutions of $q_{cc} =q_{cp} =q_{pp}$, which delineate the CP region. $\alpha^*$ is obtained from a structural percolation analysis of the periphery. In the simulations, we define a global cascade as the final fraction of active nodes exceeding $20\%$ of the entire network (see the Supplemental Material). (c)-(e) show $\rho_\infty$ as a function of $\alpha$ for three representative values of $r$. The solid lines show the results obtained using the tree-like approximation. Other parameters in the simulations are $N=10^5$, $E=5\times 10^5$, $\rho_0=10^{-5}$, $\phi=0.18$, and $\mu=0.2$. Results are averaged over $10^4$ realizations.}
\label{Fig:2}
\end{figure}

\textit{Parameter space ($r$,$\alpha$).} --- Let us first consider the case of a fixed coupling strength by keeping the parameter $\mu$ constant. Fig. \ref{Fig:2}(a) shows $\rho_\infty$ as a function of $r$ and $\alpha$. It is observed that global cascades occur within an intermediate range of $\alpha$ inside the CP region. When $\alpha$ is either too small or too large, the cascade is localized and fails to spread throughout the network. The global cascade regime is bounded by a lower (small $\alpha$) and an upper (large $\alpha$) phase boundary, corresponding to distinct phase transitions, as will be discussed below.

To intuitively understand why information is difficult to spread when $\alpha$ is large, we examine separately the roles of the core and periphery in the spreading process. For large values of $\alpha$, most links are assigned to the core, leaving the periphery sparsely connected. The high link density of the core makes it difficult for information to percolate through it, since high-degree nodes require more active neighbors to satisfy the threshold condition, exhibiting strong structural inertia (i.e., resistance to the state change). Meanwhile, the periphery is fragmented into many isolated components due to the lack of links, in which information cannot propagate effectively.

\begin{figure}[tbp]
\centering
\includegraphics[width=0.80\linewidth]{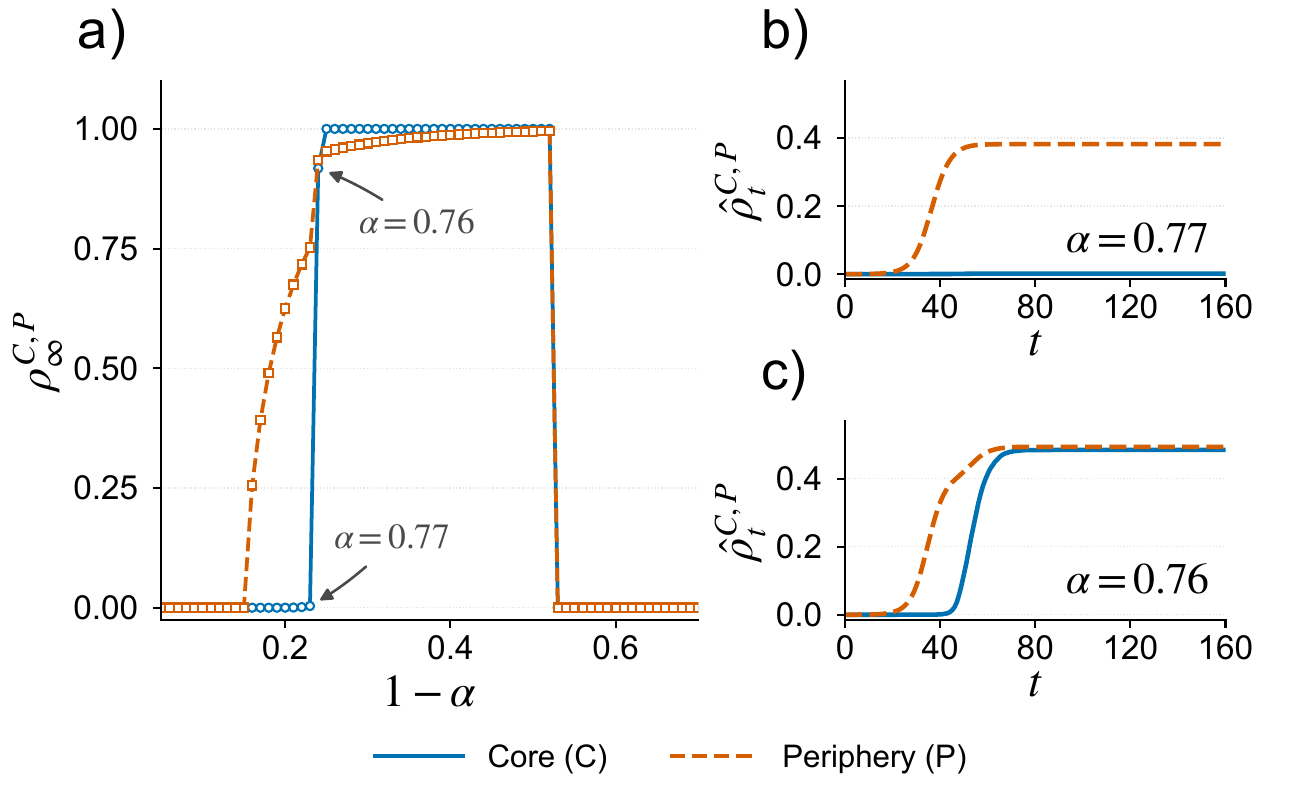}
\caption{(color online). (a) Average normalized sizes of global cascades in the core and periphery at the steady state, $\rho_\infty^C$ (circles) and $\rho_\infty^P$ (squares), versus $1-\alpha$. (b) and (c) Time evolution of the fraction of active nodes in the core ($\hat \rho^C$) and periphery ($\hat \rho^P$) at two representative values, $\alpha=0.77$ and $\alpha=0.76$. Note that $\rho_\infty^X$ is conditioned on global cascades, whereas $\hat \rho^X$ is calculated based on all cascades ($X=C,P$). Parameters in the simulations are $N=10^5$, $E=5\times 10^5$, $\phi=0.18$, $\rho_0=10^{-5}$, $\mu=0.2$, and $r=0.05$. Results are averaged over $10^4$ realizations.}
\label{Fig:3}
\end{figure}

The interplay between the structural inertia of the core and the structural constraints of the periphery can give rise to double phase transitions at the upper boundary. In particular, when $r$ is small (e.g., $r=0.05$), as $\alpha$ increases, $\rho_\infty$ undergoes a discontinuous drop at $\alpha_{c1}^U \approx 0.76$, after which it decreases continuously, reaching $\rho_\infty=0$ at the second critical point $\alpha_{c2}^U \approx 0.85$ [Fig.\ref{Fig:2}(c)], corresponding to a discontinuous-continuous type transition. To better understand this behavior, let us consider the fully core-concentrated limit (i.e., $\alpha=1$) and gradually decrease $\alpha$. In such process, a giant connected component in the periphery gradually emerges and dominates the cascade (its relatively low average degree makes it susceptible to the initial seeds), while the highly connected core remains largely inactive [Fig.\ref{Fig:3} (b)]. As $\alpha$ decreases further, the fraction of active nodes in the periphery continues to grow as the giant component expands. Consequently, the spreading dynamics become two-stage, with an initial outbreak in the periphery, followed by a delayed core activation triggered by peripheral nodes [Fig. \ref{Fig:3} (c)]. The activated core can further feed back into the periphery through the interconnecting links, activating those peripheral nodes outside the giant component that would otherwise remain inactive. The second outbreak in the periphery is reflected in the discontinuous jump in the $\rho_\infty^p(\alpha)$ curve, as shown in Fig.\ref{Fig:3} (a).

It should be emphasized that $\alpha_{c2}^U$ is not determined solely by the structural connectivity of the periphery. To illustrate this, we estimate the critical point $\alpha^*$ at which the periphery becomes fragmented. For a random network with a Poisson degree distribution, a giant component emerges when its average degree exceeds one \cite{Callaway2000}. Applying this criterion to the periphery, we obtain $\alpha^*=1-\frac{(1-r)N}{2E(1-\mu)}$, which is greater than $\alpha_{c2}^U$. This result indicates that global cascades disappear before the periphery becomes structurally fragmented. The discrepancy arises from the coupling between the two blocks: the interconnection links of peripheral nodes increase their total degree without providing a comparable increase in the number of active neighbors, making it more difficult for periphery nodes to change state.


The emergence of the double phase transition depends strongly on the relative core size $r$. As shown in Fig.\ref{Fig:2} (e), when $r$ is sufficiently large (e.g., $r=0.4$), the double transition disappears, replaced by a single discontinuous jump. This can be attributed to the dilution of the core link density for large $r$, which makes the core vulnerable to small perturbations from the periphery. As a result, the core can be activated collectively without waiting for a sufficient accumulation of active peripheral nodes. Thus, the gradual growth of the peripheral cascade is bypassed. Interestingly, before the double transition collapses into a single discontinuous transition, its type changes from discontinuous-continuous to discontinuous-discontinuous [Fig.\ref{Fig:2} (d)]. The second discontinuous transition (although not pronounced) further confirms that the cascading behavior near $\alpha_{c2}^U$ is not fully constraint to the giant component in the periphery, especially when the initial seed size is extremely small (see the Supplemental Material).

We next focus on the lower boundary. When $\alpha$ decreases, the core becomes sparser and more links are redistributed to the periphery. The resulting increase in the average degree of the periphery can induce a discontinuous phase transition from a global cascade to an absorbing state, as previously shown for random networks \cite{Watts2002}. However, if information were able to propagate through the core, global cascades could still occur even for  small $\alpha$, in which case the lower phase transition would disappear. By comparing the average degree of the core, $z_c\approx q_{cc}N_c+q_{cp}N_p$, and that of the periphery, $z_p\approx q_{cp}N_c+q_{pp}N_p$, we find that $z_c$ is consistently higher than $z_p$ (note that $q_{cc}>q_{cp}>q_{pp}$). This indicates that when $\alpha$ becomes small such that activation in the periphery is impossible, activation in the core is even less likely.

\begin{figure}[tbp]
\centering
\includegraphics[width=0.52\linewidth]{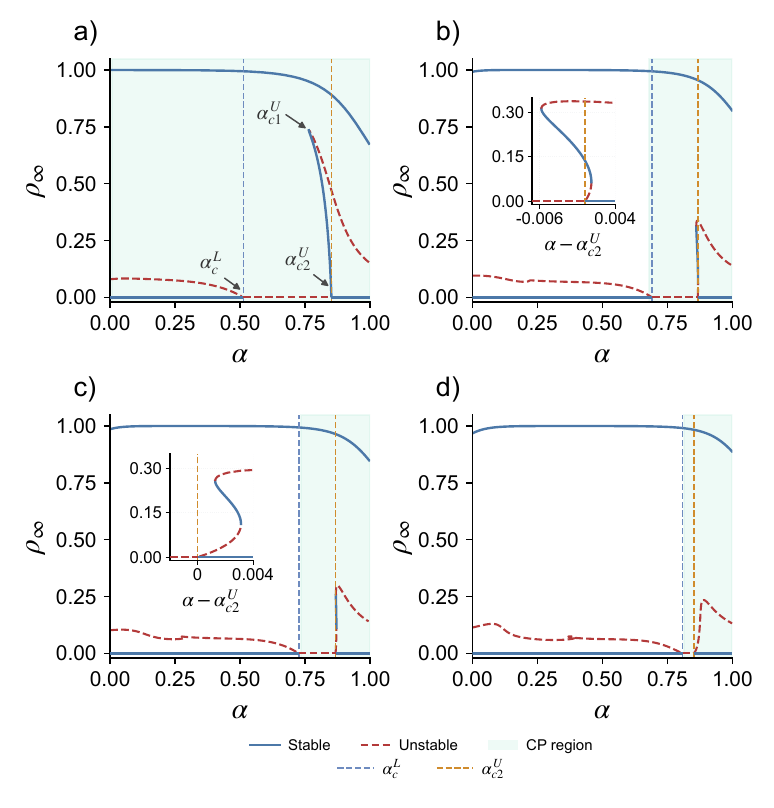}
\caption{(color online). Bifurcation diagrams for various values of $r$ when $\rho_0 \to 0$. (a) $r=0.05$, (b) $r=0.28$, (c) $r=0.32$, and (d) $r=0.4$. The vertical dashed lines are obtained by solving Eq. (\ref{eq:condition-theory}). The other parameters in the model are $N=10^5$, $E=5\times 10^5$, $\phi=0.18$, and $\mu=0.2$.}
\label{Fig:4}
\end{figure}

The emergence of different types of phase transitions in the model can be understood through a bifurcation analysis of Eqs.(\ref{eq:yn-yp1}) and (\ref{eq:yn-yp2}). Taking $\mu=0.2$ as an example, Fig. \ref{Fig:4} illustrates the stable and unstable solutions of the above equations under different values of $r$. For small $r$ [e.g., $r=0.05$, see Fig. \ref{Fig:4} (a)], at the lower transition (the left part of the bifurcation diagram), two stable branches coexist with an intermediate unstable branch for $\alpha<\alpha_c^L\approx 0.51$. As $\alpha$ approaches $\alpha_c^L$, the unstable branch collides with the lower stable branch and both disappear through a saddle-node bifurcation, leading to a discontinuous transition. At the upper transition, the stable and unstable branches exchange stability at $\alpha_{c2}^U\approx 0.85$ via a transcritical bifurcation, indicating a continuous transition. Notably, the stable branch generated by the transcritical bifurcation ($\rho_\infty>0$) collides with another unstable branch, and annihilates at $\alpha_{c1}^U\approx 0.76$ through a saddle-node bifurcation, giving rise to a discontinuous transition. Together with the continuous transition at $\alpha_{c2}^U$, this forms the observed discontinuous--continuous double phase transition. With the increase of $r$ (e.g., $r=0.28$), the transcritical bifurcation at $\alpha_{c2}^U$ disappears and is replaced by a saddle-node bifurcation, which can generate a small discontinuous jump from the lower stable branch ($\rho_\infty=0$) [Fig. \ref{Fig:4} (b)]. Further increasing $r$, the lower stable branch undergoes a large discontinuous jump to the upper stable branch ($\rho_\infty \approx 1$) at $\alpha_{c2}^U$ [Fig.\ref{Fig:4}(c) and (d)].

Furthermore, the condition for global cascades, particularly for $\alpha_{c}^L$ and $\alpha_{c2}^U$, can also be computed through a linear stability analysis. In the limit of infinitesimal initial seed density, i.e., $\rho_0^p \to 0$ (and recall that $\rho_0^c=0$), Eqs. (\ref{eq:yn-yp1}) and (\ref{eq:yn-yp2}) can be linearized around the inactive fixed point $(0,0)$, yielding a two-dimensional Jacobian matrix $\boldsymbol{J}$ (see the Supplemental Material). A global cascade occurs only if $y_n^c$ or $y_n^p$ increases with $n$, which requires the largest eigenvalue of $\boldsymbol{J}$ to exceed $1$. This condition gives  
\begin{eqnarray}\label{eq:condition-theory}
\frac{T_c\theta_c+T_p\theta_p+\sqrt{(T_c\theta_c-T_p\theta_p)^2+4T_cT_p\bar\theta_c\bar\theta_p}}{2} > 1,
\end{eqnarray}
where $\theta_c=\frac{z_c^{in}}{z_c}$, $\bar \theta_c=\frac{z_c^{out}}{z_c}$, $\theta_p=\frac{z_p^{in}}{z_p}$, $\bar\theta_p=\frac{z_p^{out}}{z_p}$, $T_c=\frac{1}{z_c}\sum_{k=1}^{\lfloor 1/\phi \rfloor} k(k-1) P_c(k)$, and $T_p=\frac{1}{z_p}\sum_{k=1}^{\lfloor 1/\phi \rfloor} k(k-1) P_p(k)$.

\textit{Parameter space ($\mu$,$\alpha$).}
In this section, we examine the $(\mu,\alpha)$ parameter space by fixing the relative core size $r$. As illustrated in Fig.\ref{Fig:2} (b), we again observe a global cascade regime within the CP region. The phase transition behavior of $\rho_\infty(\alpha)$ is qualitatively similar to that observed in the $(r,\alpha)$ parameter space. In particular, when $\mu$ is relatively small, a clear double phase transition emerges at the upper boundary. For larger $\mu$, the double phase transition becomes a single discontinuous transition. This can be attributed to the increasing number of interconnecting links, which increases the opportunity that core nodes are exposed to active peripheral nodes. As a result, the core can be activated before a sufficient number of active peripheral nodes has accumulated. A further increase in $\mu$, however, suppresses global cascading, as the periphery becomes too sparse and the core nodes become too stable to fulfill the threshold condition.


In summary, we have investigated how CP structures modulate complex spreading dynamics within the framework of the linear threshold model. The CP network is constructed based on the stochastic block model controlled by three parameters: the relative core size $r$, the core-periphery coupling $\mu$, and the intra-block link allocation to the core $\alpha$. We show that a global cascade window emerges in the parameter space $(r,\alpha)$ or $(\mu,\alpha)$, bounded by distinct types of phase transitions. For small $\alpha$, the spreading process is governed primarily by threshold dynamics within the periphery, leading to a discontinuous transition. For large $\alpha$, the periphery becomes sparsely connected and constrains the spreading, while the dense core is strongly resistant to activation. The interplay between these two effects determines the nature of the upper phase boundary. In particular, for appropriate values of $r$ (or $\mu$), the system exhibits double phase transitions via a two-stage cascading mechanism. As $r$ (or $\mu$) increases, the double transition changes type from discontinuous-continuous to discontinuous-discontinuous, and eventually disappears, replaced by a single discontinuous transition. Our model reveals a simple structural mechanism for double phase transitions and their type changes in spreading dynamics. These findings have potential implications not only for understanding information spreading in social networks, but also for related cascading phenomena in other networked systems, such as cascading failures in power grids.

\vspace{\baselineskip}
This work was supported in part by the Key R\&D Programm of Zhejiang (2024C01025), the National Key R\&D Program of China (2025YFA1510900), the Research and Development Program of Zhejiang (2026C02A1264, 2026C02A1233), the General Project of Humanities and Social Sciences Research, Ministry of Education of China (23YJA860011), the Fundamental Research Funds for the Central Universities (1243200012), and the Guangdong Philosophy and Social Science Foundation Regular Project (GD24XXW02).



\clearpage

\clearpage
\setcounter{equation}{0}
\setcounter{figure}{0}
\setcounter{table}{0}
\renewcommand{\theequation}{S\arabic{equation}}
\renewcommand{\thefigure}{S\arabic{figure}}
\renewcommand{\thetable}{S\arabic{table}}
\begin{center}
\begin{minipage}{\textwidth}
\centering
{\bfseries Supplemental Material for\\[0.4\baselineskip]
Core-periphery structure induces double phase transitions in information cascades\par}
\vspace{0.8\baselineskip}
Jiaqi Nie\textsuperscript{1,2,3,*},
Xincheng Shu\textsuperscript{4,5,*},
Yong Min\textsuperscript{4,5},
Qi Xuan\textsuperscript{1,2}, and
Zhongyuan Ruan\textsuperscript{1,\ensuremath{\dagger}}\par
\vspace{0.5\baselineskip}
{\small\itshape
\textsuperscript{1}Institute of Cyberspace Security, Zhejiang University of Technology, Hangzhou 310023, China\par
\textsuperscript{2}Binjiang Institute of Artificial Intelligence, Zhejiang University of Technology, Hangzhou 310056, China\par
\textsuperscript{3}Department of Network and Data Science, Central European University, Vienna, 1100, Austria\par
\textsuperscript{4}Computational Communication Research Center, Beijing Normal University, Zhuhai, 519087, China\par
\textsuperscript{5}School of Journalism and Communication, Beijing Normal University, Beijing, 100875, China\par}
\vspace{0.5\baselineskip}
{\small
\textsuperscript{*}These authors contributed equally to this work.\par
\textsuperscript{\ensuremath{\dagger}}Corresponding author: \texttt{zyuan.ruan@gmail.com}\par}
\end{minipage}
\end{center}
\vspace{0.5\baselineskip}

\subsection*{S1. Metrics for global cascades} 
In the simulations, we define a global cascade as the final fraction of active nodes $f_\infty$ exceeding $20\%$ of the entire network. The simulation results are robust to different choices of the criteria. The average size of global cascades $\rho_\infty$ is thus calculated as 
\begin{equation}\label{supp:eq:rho}
\rho_\infty =
\begin{cases}
  \frac{1}{R}\sum_{i=1}^R f_\infty^i , & f_\infty^i > 20\%, \\
  0, & otherwise,
\end{cases}
\end{equation}
where $R$ is the number of realizations that trigger a global cascade, and $f_\infty^i$ denotes the final fraction of active nodes in realization $i$. 
Note that $\rho_\infty$ is defined as the average cascade size conditioned on the occurrence of a global cascade. This definition is consistent with the theoretical analysis based on the tree-like approximation when $\rho_0 \to 0$ \cite{Gleeson:2008}. Therefore, the simulated $\rho_\infty$ can be directly compared to the theoretical predictions. For a finite seed fraction $\rho_0$ (i.e., $\rho_0$ does not vanish as $N\to \infty$), the tree-like theoretical predictions can be directly compared to the average cascade size over all realizations, which we denote by $\hat \rho_\infty$.


\subsection*{S2. Calculations for $\bar y_n^{c}$ and $\bar y_n^{p}$} 
We adopt the tree-like approximation developed by Gleeson to analyze our model \cite{Gleeson:2007}. Specifically, we approximate the original network by an infinite-depth tree, with the levels labeled from $n=0$ at the bottom to $n=\infty$ at the top. A key assumption of this approach is that node states update sequentially from the bottom to the top: a node at level $n$ can only be influenced by child nodes at level $n-1$. We define $y_n^c$ (and $y_n^p$) as the probability that a randomly selected core (periphery) node at level $n$ is active, conditioned on its parent node at level $n+1$ being inactive (denoted as condition $C$), i.e., 
\begin{equation}\label{supp:eq:yn}
y_n^{c}=\frac{\#\{\text{active core nodes at level }n \mid C\} }{\#\{\text{core nodes at level }n \mid C\} }.
\end{equation}
The quantity $\bar y_n^{c}$ (respectively, $\bar y_n^{p}$) is defined as the probability that a randomly chosen child node at level $n$ of a core (periphery) node at level $n+1$ is active. It can be computed as 
\begin{equation}\label{supp:eq:bar-yn}
\begin{aligned}
\bar y_n^{c} &= \frac{z_{c}^{in}}{z_{c}} y_n^{c} +\frac{z_{c}^{out}}{z_{c}} y_n^{p},\\
\bar y_n^{p} &= \frac{z_{p}^{in}}{z_{p}} y_n^{p} +\frac{z_{p}^{out}}{z_{p}} y_n^{c}.
\end{aligned}
\end{equation}
Here, $z_{c}=z_{c}^{in}+z_{c}^{out}$ is the average degree of a core node, where $z_{c}^{in}$ and $z_{c}^{out}$ denote the average numbers of links connecting to the core and periphery, respectively. Accordingly, $z_{c}^{in}/z_{c}$ and $z_{c}^{out}/z_{c}$ represent the probabilities that a randomly selected child of a core node belongs to the core and periphery, respectively. Similarly, $z_{p}=z_{p}^{in}+z_{p}^{out}$, with $z_{p}^{in}/z_{p}$ and $z_{p}^{out}/z_{p}$ denoting the corresponding probabilities for a periphery node. In our model, these quantities can be readily obtained as 
\begin{equation}\label{supp:eq:zs}
\begin{aligned}
z_c^{in} &= \frac{2(1-\mu)\alpha E}{rN}, &\qquad z_c^{out} &= \frac{\mu E}{rN},\\
z_p^{in} &= \frac{2(1-\mu)(1-\alpha)E}{(1-r)N}, &\qquad z_p^{out} &= \frac{\mu E}{(1-r)N}.
\end{aligned}
\end{equation}
Inserting the above relations into Eq. (\ref{supp:eq:bar-yn}), we have
\begin{equation}\label{supp:eq:bar-yn-solution}
\begin{aligned}
\bar y_n^{c} &= \frac{2(1-\mu)\alpha}{2(1-\mu)\alpha+\mu} y_n^{c} + \frac{\mu}{2(1-\mu)\alpha+\mu} y_n^{p}, \\
\bar y_n^{p} &= \frac{\mu}{2(1-\mu)(1-\alpha)+\mu} y_n^{c} +  \frac{2(1-\mu)(1-\alpha)}{2(1-\mu)(1-\alpha)+\mu} y_n^{p}.
\end{aligned}
\end{equation}

\subsection*{S3. Linear stability analysis of the system} 
We start from the following recursive equations:
\begin{equation}\label{supp:eq:yn-yp}
\begin{aligned}
y_{n+1}^{c}&=\rho_0^{c}+(1-\rho_0^{c})F(y_n^c,y_n^p), \\
y_{n+1}^{p}&=\rho_0^{p}+(1-\rho_0^{p})G(y_n^c,y_n^p), 
\end{aligned}
\end{equation}
which can be interpreted as follows. The probability that a randomly selected core (or periphery) node at level $n+1$ is active consists of two contributions: (i). the node is an initial seed, with probability $\rho_0^{c}$ ($\rho_0^{p}$), or (ii). the node is initially inactive but subsequently becomes active through the influence of its child nodes at level $n$, which occurs with probability $F(y_n^c,y_n^p)$ (or $G(y_n^c,y_n^p)$). The functions $F(y_n^c,y_n^p)$ and $G(y_n^c,y_n^p)$ are given by
\begin{equation}\label{supp:eq:F-G}
\begin{aligned}
F&=\sum_k \frac{kP_{c}(k)}{z_c} \sum_{m=\lceil \phi k\rceil}^{k-1}  \begin{pmatrix} k-1\\  m \end{pmatrix} (\bar y_n^{c})^m (1-\bar y_n^{c})^{k-1-m}, \\ 
G&=\sum_k \frac{kP_{p}(k)}{z_p} \sum_{m=\lceil \phi k\rceil}^{k-1}  \begin{pmatrix} k-1\\  m \end{pmatrix} (\bar y_n^{p})^m (1-\bar y_n^{p})^{k-1-m}, 
\end{aligned}
\end{equation}
where $P_{c}(k)$ and $P_{p}(k)$ are the degree distributions of the core and periphery, respectively, and $\phi$ is the adoption threshold. In the limit $\rho_0^p \to 0$ and $\rho_0^c=0$, we linearize Eq. (\ref{supp:eq:yn-yp}) around the fixed point $(0,0)$, yielding
\begin{equation}\label{supp:eq:matrix_form1}
\begin{aligned}
\begin{pmatrix} y_{n+1}^c\\  y_{n+1}^p \end{pmatrix} &= \begin{bmatrix}  \frac{\partial F}{\partial y_n^c} &  \frac{\partial F}{\partial y_n^p} \\[8pt]  \frac{\partial G}{\partial y_n^c} &  \frac{\partial G}{\partial y_n^p} \end{bmatrix}_{0,0} \begin{pmatrix} y_{n}^c\\  y_{n}^p \end{pmatrix} \equiv \boldsymbol{J}  \begin{pmatrix} y_{n}^c\\  y_{n}^p \end{pmatrix},
\end{aligned}
\end{equation}
where $\boldsymbol{J}$ is the Jacobian matrix evaluated at the origin. After some algebra (by calculating the partial derivatives), we obtain 
\begin{equation}\label{supp:eq:J-matrix}
\begin{aligned}
\boldsymbol{J} = \begin{pmatrix} T_c \theta_c & T_c \bar \theta_c \\  T_p \bar \theta_p & T_p  \theta_p \end{pmatrix},
\end{aligned}
\end{equation}
with 
\begin{equation}\label{supp:eq:elements}
\begin{aligned}
\theta_c&=\frac{z_c^{in}}{z_c}= \frac{2(1-\mu)\alpha}{2(1-\mu)\alpha+\mu}, &\qquad
\bar \theta_c&=\frac{z_c^{out}}{z_c}=\frac{\mu}{2(1-\mu)\alpha+\mu},\\
\theta_p&=\frac{z_p^{in}}{z_p}=\frac{2(1-\mu)(1-\alpha)}{2(1-\mu)(1-\alpha)+\mu}, &\qquad
\bar\theta_p&=\frac{z_p^{out}}{z_p}=\frac{\mu}{2(1-\mu)(1-\alpha)+\mu},\\
T_c&=\frac{1}{z_c}\sum_{k=1}^{\lfloor 1/\phi \rfloor} k(k-1) P_c(k), &\qquad
T_p&=\frac{1}{z_p}\sum_{k=1}^{\lfloor 1/\phi \rfloor} k(k-1) P_p(k).
\end{aligned}
\end{equation}

A global cascade occurs if a small fraction of initial seeds ($\rho_0\ll 1$) results in a large fraction of active nodes, meaning that $y_n^c$ or $y_n^p$ should increase with $n$. This requires that the largest eigenvalue of the Jacobian matrix to exceed one, i.e., $\lambda_{max}(\boldsymbol{J})>1$. Thus, the condition for a global cascade is
\begin{equation}\label{supp:eq:condition}
\frac{T_c\theta_c+T_p\theta_p+\sqrt{(T_c\theta_c-T_p\theta_p)^2+4T_cT_p\bar\theta_c\bar\theta_p}}{2} > 1.
\end{equation}

\subsection*{S4. Influence of initial seed size} 
\begin{figure}[!htbp]
\centering
\includegraphics[width=0.52\linewidth]{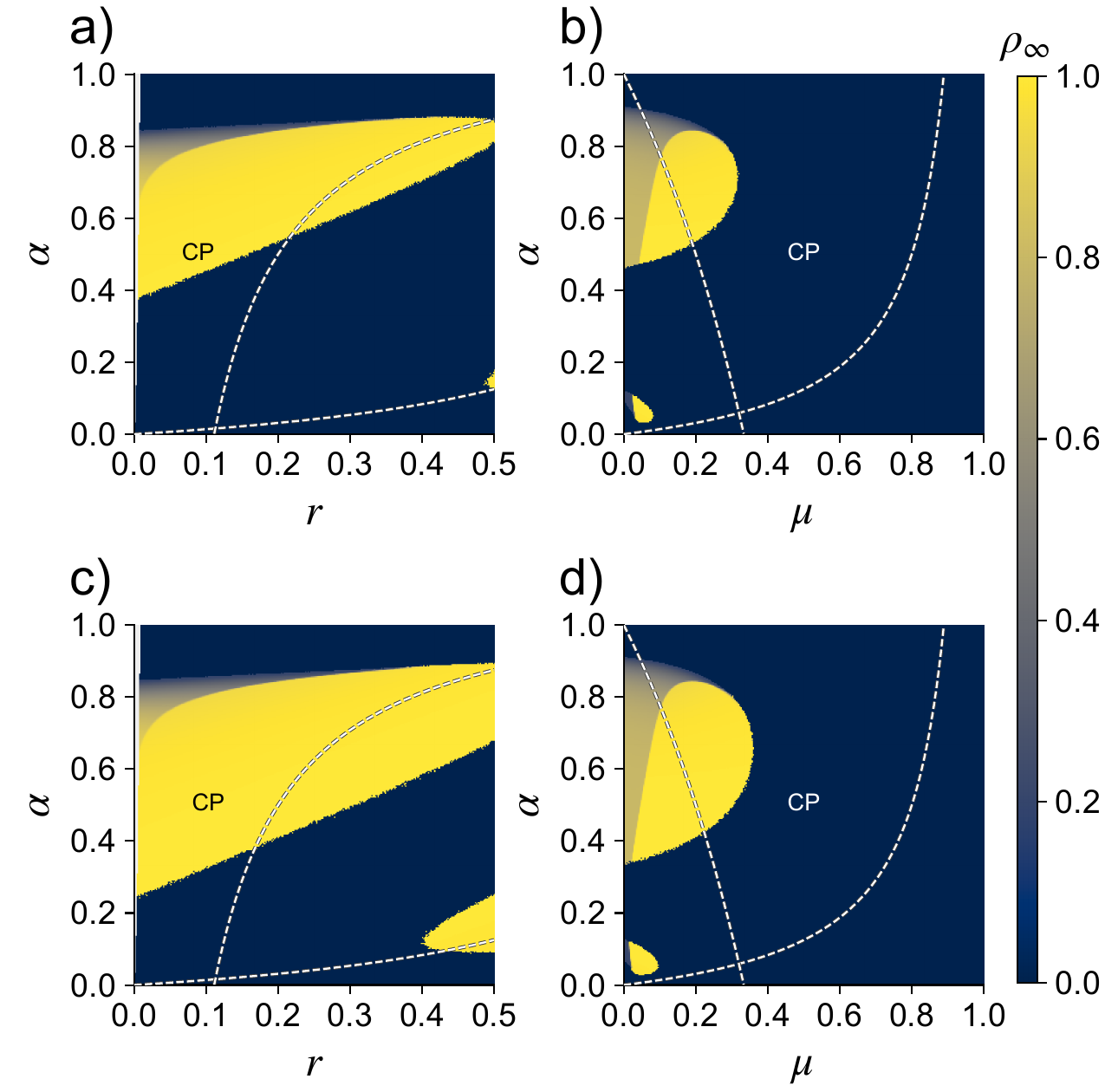}
\caption{(color online). Phase diagrams for different initial seed sizes. (a) $\rho_0=10^{-3}$ and $\mu=0.2$, (b) $\rho_0=10^{-3}$ and $r=0.2$, (c) $\rho_0=5\times 10^{-3}$ and $\mu=0.2$, (d) $\rho_0=5\times 10^{-3}$ and $r=0.2$. The other parameters in the simulations are $N=10^5$, $E=5\times 10^5$, and $\phi=0.18$. Results are averaged over $10^4$ realizations.} 
\label{Fig:S1}
\end{figure}

\begin{figure}[!htbp]
\centering
\includegraphics[width=0.70\linewidth]{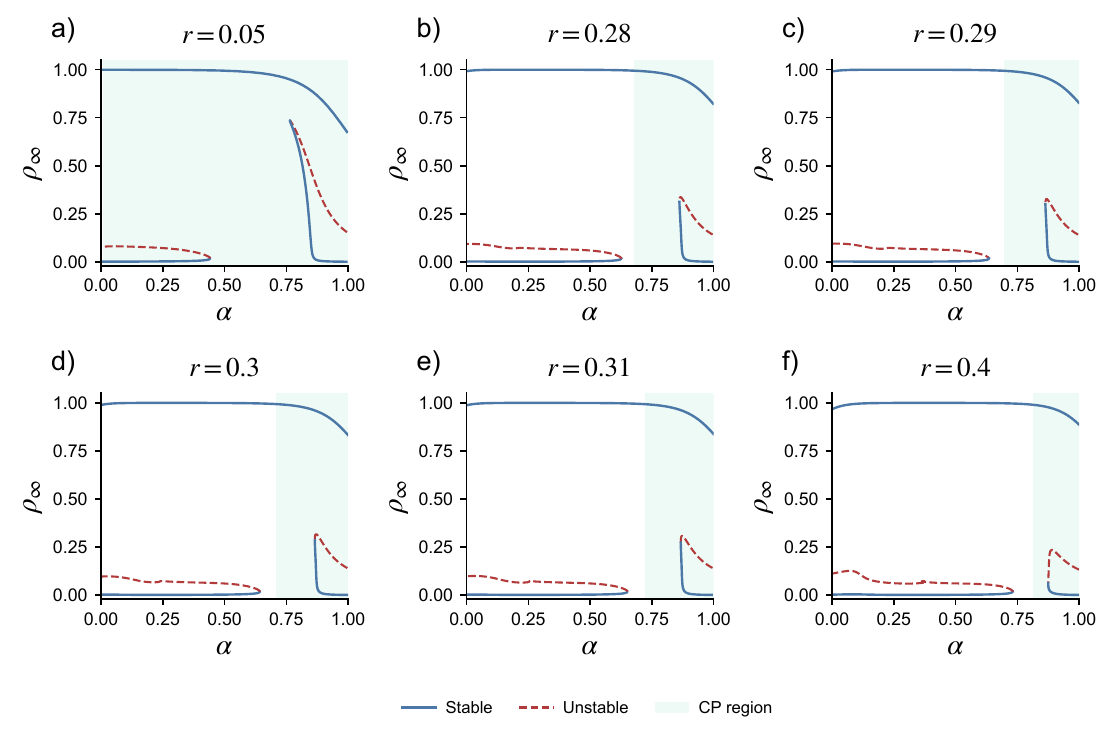}
\caption{(color online). Bifurcation diagrams for various values of $r$ when $\rho_0=10^{-3}$. (a) $r=0.05$, (b) $r=0.28$, (c) $r=0.29$, (d) $r=0.30$, (e) $r=0.31$, and (f) $r=0.4$. The other parameters in the model are $N=10^5$, $E=5\times 10^5$, $\mu=0.2$, and $\phi=0.18$.}
\label{Fig:S2}
\end{figure}

Figure \ref{Fig:S1} shows the phase diagrams for different initial seed sizes. As expected, the global cascade region expands as $\rho_0$ increases, since a larger number of initial seeds increases the change that nodes encounter active neighbors, thereby facilitating spreading. Despite this difference, the double transitions observed in the main text are robust for different initial seed sizes. However, for larger $\rho_0$, the change in the double-transition type disappears --- the double transition remains of the discontinuous-continuous type as $r$ increases (see Fig. \ref{Fig:S2}). 

To understand this behavior, let us decrease $\alpha$ from $1$ and focus on the regime near $\alpha_{2c}^U$. During this process, a giant connected component (GCC) in the periphery gradually emerges. For relatively large $\rho_0$, the GCC can be activated with high probability. The spreading therefore is mainly constrained by the size of the GCC, which grows continuously with $\alpha$, resulting in a continuous transition. In contrast, for an extremely small initial seed size, the emergence of a GCC in the periphery is not sufficient to trigger a global cascade. Due to the presence of additional links connecting to the core, peripheral nodes in the GCC may fail to satisfy their threshold conditions (the number of initially active nodes is too small). In this case, the threshold dynamics plays an essential role and can produce a discontinuous transition at $\alpha_{2c}^U$. Nevertheless, the cascade size is still restricted by the size of the GCC. 
 
\clearpage

\renewcommand{\lastbiblabel}{LastSupplementalBibItem}


\begin{thebibliography}{99}

\bibitem{Rombach2017}
P. Rombach, M. A. Porter, J. H. Fowler, and P. J. Mucha, SIAM Review {\bf 59}, 619 (2017).

\bibitem{Newman2006}
M.E.J. Newman, Proc. Natl. Acad. Sci. USA {\bf 103},  8577 (2006).

\bibitem{Nematzadeh2014}
A. Nematzadeh, E. Ferrara, A. Flammini, and Y. -Y. Ahn, Phys. Rev. Lett. {\bf 113}, 088701 (2014).

\bibitem{Verma2014}
T. Verma, N. A. Araújo, and H. J. Herrmann, Sci. Rep. {\bf 4}, 5638 (2014).

\bibitem{Fagiolo2010}
G. Fagiolo, J. Reyes, and S. Schiavo, J. Evol. Econ. {\bf 20}, 479-514 (2010).

\bibitem{Masuda2006}
N. Masuda and N. Konno, PLoS One {\bf 28}, 297-309 (2006).

\bibitem{Anastasiou2019}
D. Anastasiou, H. Louri, and M. Tsionas, Int. J. Financ. Econ. {\bf 24}, 97 (2019).

\bibitem{Lee2014}
S. H. Lee, M. Cucuringu, M. A. Porter, Phys. Rev. E {\bf 89}, 032810 (2014).

\bibitem{Borgatti2000}
S. P. Borgatti and M. G. Everett, Social networks {\bf 21}, 375-395 (2000).

\bibitem{Zhang2015}
X. Zhang, T. Martin, M. E. J. Newman, Phys. Rev. E {\bf 91}, 032803 (2015).

\bibitem{Rossa2013}
F. D. Rossa, F. Dercole, and C. Piccardi, Sci. Rep. {\bf 3}, 1467 (2013).

\bibitem{Holme2005}
P. Holme, Phys. Rev. E {\bf 72}, 046111 (2005).

\bibitem{Battiston2018}
F. Battiston, J. Guillon, M. Chavez, V. Latora, and F. D. V. Fallani, J. R. Soc. Interface {\bf 15}, 20180514 (2018).

\bibitem{Nie2025}
J. Nie, Q. Xuan, D. Gao, and Z. Ruan, Physica A {\bf 669}, 130618 (2025).

\bibitem{Bergermann2025}
K. Bergermann and F. Tudisco, Phys. Rev. Lett. {\bf 135}, 087402 (2025).

\bibitem{Verma2016}
T. Verma, F. Russmann, N. A. M. Araújo, J. Nagler, H. J. Herrmann, Nat. Comm. {\bf 7}, 10441 (2016).

\bibitem{Colomer2014}
P. Colomer-de-Simón and M. Boguñá, Phys. Rev. X {\bf 4}, 041020 (2014).

\bibitem{Kitsak2010}
M. Kitsak, L. K. Gallos, S. Havlin, F. Liljeros, L. Muchnik, H. E. Stanley, and H. A. Makse, Nat. Phys. {\bf 6}, 888 (2010).

\bibitem{Malliaros2016}
F. D. Malliaros, M. E. G. Rossi, and M. Vazirgiannis, Sci. Rep. {\bf 6}, 19307 (2016).

\bibitem{Nadini2020}
M. Nadini, L. Zino, A. Rizzo, and M. Porfiri, Appl. Netw. Sci. {\bf 5}, 68 (2020).

\bibitem{Huang2019}
J. Huang, and J. Huang, Physica A {\bf 528}, 121452 (2019).

\bibitem{Granovetter1983}
M. Granovetter and R. Soong, J. Math. Sociol {\bf 9}, 165 (1983).

\bibitem{Karsai2014}
M. Karsai, G. Iñiguez, K. Kaski, and J. Kertész, J. R. Soc. Interface {\bf 11}, 20140694 (2014).

\bibitem{Ruan2015}
Z. Ruan, G. Iñiguez, M. Karsai, and J. Kertész, Phys. Rev. Lett. {\bf 115}, 218702 (2015).

\bibitem{Watts2002}
D. J. Watts, Proc. Natl. Acad. Sci. U.S.A. 99, 5766 (2002).

\bibitem{Payne2009}
J. L. Payne, P. S. Dodds, and M. J. Eppstein, Phys. Rev. E {\bf 80}, 026125 (2009).

\bibitem{Yagan2012}
O. Yagan and V. Gligor, Phys. Rev. E {\bf 86}, 036103 (2012).

\bibitem{Galstyan2007}
A. Galstyan and P. Cohen, Phys. Rev. E 75, 036109 (2007).

\bibitem{Curato2016}
G. Curato and F. Lillo, Phys. Rev. E {\bf 94}, 032310 (2016).

\bibitem{Fu2025}
G. Fu and H. Liang, Mathematics {\bf 13}, 3048 (2025).

\bibitem{Min2018}
B. Min and M. San Miguel, Sci. Rep. {\bf 8}, 10422 (2018).

\bibitem{Shu2024}
X. Shu, M. Yang, Q. Xuan, and Z. Ruan, Chaos Solitons Fractals {\bf 189}, 115687 (2024).

\bibitem{Yang2023}
Z. Yang, J. Wu, J. He, K. Xu, and M. Zheng, Chaos Solitons Fractals {\bf 171}, 113487 (2023).

\bibitem{Gao2015}
J. Gao, T. Zhou, and Y. Hu, Sci. Rep. {\bf 5}, 14662 (2015).

\bibitem{Wang2018}
W. Wang, Z. Wang, and S. Cai, Phys. Rev. E {\bf 98}, 052312 (2018).

\bibitem{Gleeson2007}
J. P. Gleeson and D. J. Cahalane, Phys. Rev. E {\bf 75}, 056103 (2007).

\bibitem{Gleeson2008}
J. P. Gleeson, Phys. Rev. E {\bf 77}, 046117 (2008).

\bibitem{Callaway2000}
D. S. Callaway, M. E. J. Newman, S. H. Strogatz, and D. J. Watts, Phys. Rev. Lett. {\bf 85}, 5468 (2000).

\end{thebibliography}

\begin{thebibliography}{9}
\bibitem{Gleeson:2008}
J. P. Gleeson, Cascades on correlated and modular random networks, Phys. Rev. E 77, 046117 (2008).

\bibitem{Gleeson:2007}
J. P. Gleeson and D. J. Cahalane, Seed size strongly affects cascades on random networks, Phys. Rev. E 75, 056103 (2007).
\end{thebibliography}
\end{document}